\documentclass[twocolumn,showpacs,unsortedaddress,superscriptaddress,showkeys,preprintnumbers,letterpaper]{revtex4-1}
\usepackage{acronym}
\usepackage{amsmath}
\usepackage{amssymb}
\usepackage{graphics}
\usepackage{subfigure}
\usepackage[usenames]{color}
\usepackage[normalem]{ulem}
\usepackage[bookmarksnumbered, bookmarksopen, breaklinks, colorlinks]{hyperref}
\pdfoutput=1

\newcommand{\LIGOCaltech}{LIGO Laboratory, California Institute of Technology, 
Pasadena, CA 91125, USA}
\newcommand{\TAPIR}{Theoretical Astrophysics, California Institute of
Technology, Pasadena, CA 91125, USA}
\newcommand{\CITA}{Canadian Institute for Theoretical Astrophysics, 60 St.
George Street, University of Toronto, Toronto, ON M5S 3H8, Canada}
\newcommand{\Perimeter}{Perimeter Institute for Theoretical Physics, Waterloo,
Ontario N2L 2Y5, Canada}
\newcommand{\AEI}{Albert-Einstein-Institut, Max-Planck-Institut f\"ur
Gravitationsphysik, D-30167 Hannover, Germany}
\newcommand{\Leibniz}{Leibniz Universit\"at Hannover, D-30167 Hannover, Germany}

\newcommand{\bankpatch}{\mathcal{P}}
\newcommand{\mat}[1]{{\bf #1}}

\newcommand{\T}{{\mathrm{T}}}

\begin{document}


\title{Efficiently enclosing the compact binary parameter space by singular-value decomposition}

\author{Kipp~Cannon}
\email{kipp.cannon@ligo.org}
\affiliation{\LIGOCaltech}
\affiliation{\CITA}

\author{Chad~Hanna}  
\email{chad.hanna@ligo.org}
\affiliation{\LIGOCaltech}
\affiliation{\Perimeter}

\author{Drew~Keppel}  
\email{drew.keppel@ligo.org}
\affiliation{\LIGOCaltech}
\affiliation{\TAPIR}
\affiliation{\AEI}
\affiliation{\Leibniz}

\begin{abstract}

Gravitational-wave searches for the merger of compact binaries use
matched-filtering as the method of detecting signals and estimating parameters.
Such searches construct a fine mesh of filters covering a signal parameter
space at high density.  Previously it has been shown that singular value
decomposition can reduce the effective number of filters required to search the
data.  Here we study how the basis provided by the singular value decomposition
changes dimension as a function of template bank density.  We will demonstrate
that it is sufficient to use the basis provided by the singular value
decomposition of a low density bank to accurately reconstruct arbitrary points
within the boundaries of the template bank.  Since this technique is purely
numerical it may have applications to interpolating the space of numerical
relativity waveforms.

\end{abstract}

\maketitle
\acrodef{BNS}{binary neutron star}
\acrodef{CBC}{compact binary coalescence}
\acrodef{GW}{gravitational-wave}
\acrodef{PN}{post-Newtonian}
\acrodef{ROC}{Receiver Operator Characteristic}
\acrodef{SNR}{signal-to-noise ratio}
\acrodef{SPA}{stationary phase approximation}
\acrodef{SVD}{singular value decomposition}

\section{Introduction}

Several broadband laser interferometer \ac{GW} detectors are operating at high
sensitivities and will continue to improve over the next decade~\cite{advLIGO,
advVirgo2005, GEO2010, Virgo2008, LIGO2009}.  As detectors improve it is
increasingly likely that \ac{GW} astronomers will observe gravitational
radiation emitted from the coalescence of compact binary systems involving
neutron stars and or stellar mass black holes~\cite{rates2010}.

Because \ac{CBC} waveforms are well modeled, \ac{GW} searches for such signals
are conducted by matched filtering the detectors' data with banks of template
waveforms, chosen to adequately cover a region of the signal parameter
space~\cite{Owen:1995tm}.  For \ac{GW} signals from the merger of compact
objects with negligible spin, this parameter space is defined by functions of
the masses of the two objects.  To search for signals within this parameter
space, a bank of templates is constructed to sample the parameter space
sufficiently densely such that there is minimal loss of \ac{SNR}.
Traditionally, template banks used to search this two-dimensional signal
parameter space have been constructed using the $(A_2)^*$ lattice
\cite{Cokelaer2007}, referred to as ``hexagonally-placed" template banks.  This
problem becomes more difficult in higher dimensions, where other types of
template placement algorithms have recently been investigated \cite{Harry2009,
Manca2010, Messenger2009, Prix2007}.

In \cite{Cannon2010} the \ac{SVD} was applied to \ac{CBC} waveforms to show
how hexagonally-placed template banks with $M$ templates could be
implemented with $N' \ll 2 M$ filters ($2M$ being the nominal number of
filters required for the $M$ 2-phase templates).  This was achieved by
truncating the \ac{SVD} of the matrix consisting of the time-series of the
template waveforms.  Here we demonstrate that the bases identified by the
\ac{SVD} is effective at spanning the space of all \ac{CBC} waveforms
within the region of parameter space sampled by the original bank.  We
find that the \ac{SVD} of a low-density bank provides a basis suitable
for constructing all the waveforms from a higher-density bank, even
waveforms at arbitrary locations within that region of parameter space.

This paper is organized as follows. Sec.~\ref{sec:embed} describes how we
apply the \ac{SVD} to approximately embed the signal manifold in a vector
space. Sec.~\ref{sec:recon} tests this embedding by reconstructing various
points in the manifold. Finally, Sec.~\ref{sec:disc} discusses possible
applications of this technique.

\section{Enclosing the signal space with Singular Value Decomposition}
\label{sec:embed}

In this section we explore how the number of basis vectors required to
reconstruct a template bank scales with the initial density of the template
bank.  We define a template bank of signal waveforms covering a patch
$\bankpatch$ of the signal manifold, which is used to test for the presence and
strength of signals from $\bankpatch$ in the detectors' data.  We construct a
signal matrix in the same manner as \cite{Cannon2010}.  Specifically, we create
a real-valued matrix $\mat{H}$ by alternately filling its rows with the real
and imaginary parts (cosine and sine) of the template waveform time series from
a \ac{CBC} template bank covering $\bankpatch$, $\mat{H} = \{H_{\alpha j}\} =
\{ \Re\vec{h}_{1}, \Im\vec{h}_{1}, \Re\vec{h}_{2}, \Im\vec{h}_{2}, ...,
\Re\vec{h}_{M}, \Im\vec{h}_{M}\}^\mathrm{T}$.

As in \cite{Cannon2010}, we constructed the template matrix with chirp masses
$M_c = M \eta^{5/6}$, where $M = m_1 + m_2$ is the total mass and $\eta = m_1
m_2 / M^2$ is the symmetric mass ratio, of $1.125 M_{\odot} \le M_c < 1.240
M_{\odot}$ and component masses of $1 M_{\odot} \le m_1,m_2 < 3 M_{\odot}$.
Template banks covering this region are created using template placement
algorithms of the LIGO Algorithms Library~\cite{LAL}. Template placement is
done in the $(\tau_0, \tau_3)$ plane, where $\tau_0$ and $\tau_3$ are defined
as
\begin{eqnarray} &\tau_0 = \frac{5}{256} \left( \pi f_0 \right)^{-8/3}
M_c^{-5/3}, \\ &\tau_3 = \frac{\pi}{8} \left( \pi f_0 \right)^{-5/3} M_c^{-2/3}
\eta^{-3/5}, \end{eqnarray}
and where $f_0$ is some fiducial frequency, which we choose to be $f_0 = 60$
Hz.

The non-spinning waveforms for each template are produced to 3.5 \ac{PN} order,
sampled at 2048 Hz, up to the Nyquist frequency of 1024 Hz.  The last 10
seconds of each waveform, whitened with the initial LIGO amplitude spectral
density, are used to construct $\mat{H}$. The \ac{SVD} is then applied to
$\mat{H}$, decomposing the matrix into two unitary matrices, $\mat{V}$ and
$\mat{U}$, and a diagonal matrix $\mat{\Sigma}$
\begin{equation} \mat{H} = \mat{V}\mat{\Sigma}\mat{U}^\T, \end{equation}
where $\mat{U}$ is a matrix composed of basis vectors (i.e., unit-norm
time-series vectors), $\mat{V}$ is a matrix composed of reconstruction
coefficients, and $\mat{\Sigma}$ is a matrix containing the singular values of
$\mat{H}$.

In \cite{Cannon2010}, it was demonstrated that truncating the reconstruction of
$\mat{H}$ to use only the $N'$ basis vectors with the largest singular values
results in an average fractional \ac{SNR} loss $\left\langle \delta\rho / \rho
\right\rangle$ proportional to the sum of the discarded singular values
squared. In this investigation, we truncate these reconstruction matrices at
$\left\langle \delta\rho / \rho \right\rangle = 10^{-7}$.  This corresponds
roughly to the truncation error of IEEE 754 32-bit floating-point numbers.

\begin{figure}
\center{\resizebox{\linewidth}{!}{\includegraphics{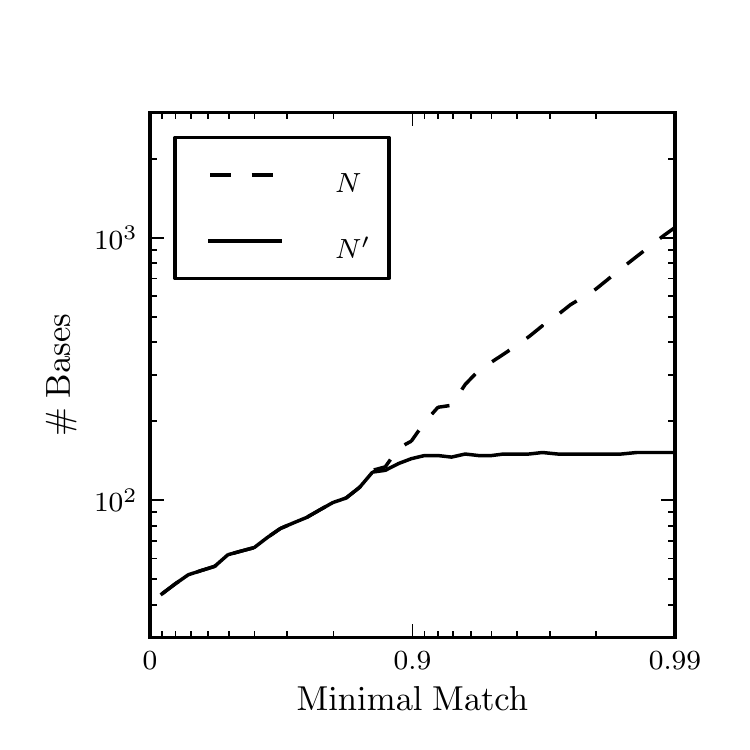}}}
\caption{The number of filters as a function of minimal match,
which increases with the density of the template bank.  The total number of
filters in the template bank, $N$, is shown by the dashed line.  The number of
filters needed to reconstruct the template matrix such that $\langle \delta
\rho / \rho \rangle = 10^{-7}$, $N'$, is shown by the solid line. We find that
the number filters needed to reconstruct $\mat{H}$ saturates when the minimal
match reaches $\sim89.9\%$.}
\label{fig:basesvsden}
\end{figure}

We explore how the number of basis vectors changes as the number of rows in
$\mat{H}$ is increased by generating template banks for $\bankpatch$ with
increasing density (i.e., increasing minimal match).  We confirmed that the
number of basis vectors required to reconstruct $\mat{H}$ saturates at a
particular value of minimal match.  Fig.~\ref{fig:basesvsden} shows that as the
minimal match of the template bank is increased, resulting in denser samplings
of $\bankpatch$, the number of basis vectors needed to reconstruct $\mat{H}$ to
the required accuracy saturates around a minimal match of $\sim89.9\%$.  This
indicates that $\bankpatch$ is able to be embedded---to an accuracy of 1 part
in $10^7$---in a vector space consisting of $\sim150$ dimensions.

In the next section we will demonstrate how the basis waveforms identified by
the coarsely sampled bank can be used to reconstruct templates at arbitrary
points on the signal manifold.

\section{Efficient reconstruction of waveforms in the manifold}
\label{sec:recon}

In order to determine how well these waveforms can be reconstructed, we compute
a quantity called the average fractional \ac{SNR} loss $\delta\rho_\alpha /
\rho_\alpha$. This quantity can be thought of as the mismatch between the
original waveform $\vec{h}_{\alpha}$ and the reconstructed waveform
$\vec{h}'_{\alpha}$, averaged over the phase angle.  It tells us how far the
reconstructed waveform is from the original waveform. As in Eq.~(25) of
\cite{Cannon2010}, $\delta\rho_\alpha / \rho_\alpha$ is given in terms of
\ac{SVD} quantities as
\begin{equation} \label{eqn:fracsnrlosseqn}
\frac{\delta\rho_\alpha}{\rho_\alpha} = \frac{1}{4}\sum_{\mu=N'+1}^{N}
\left(v_{(2\alpha-1)\mu}^2 + v_{(2\alpha)\mu}^2 \right)\sigma_{\mu}^2,
\end{equation}
where $v_{(2\alpha-1)\mu}$ and $v_{(2\alpha)\mu}$ are the reconstruction
coefficients for the real and imaginary parts, respectively, of the $\alpha$th
waveform associated with the $\mu$th basis vector and are elements of
$\mat{V}$, $\sigma_\mu$ is the $\mu$th element of $\mat{\Sigma}$, and the sum
is over the truncated terms of $\mat{V}$ and $\mat{\Sigma}$.

\begin{figure}
\center{\resizebox{\linewidth}{!}{\includegraphics{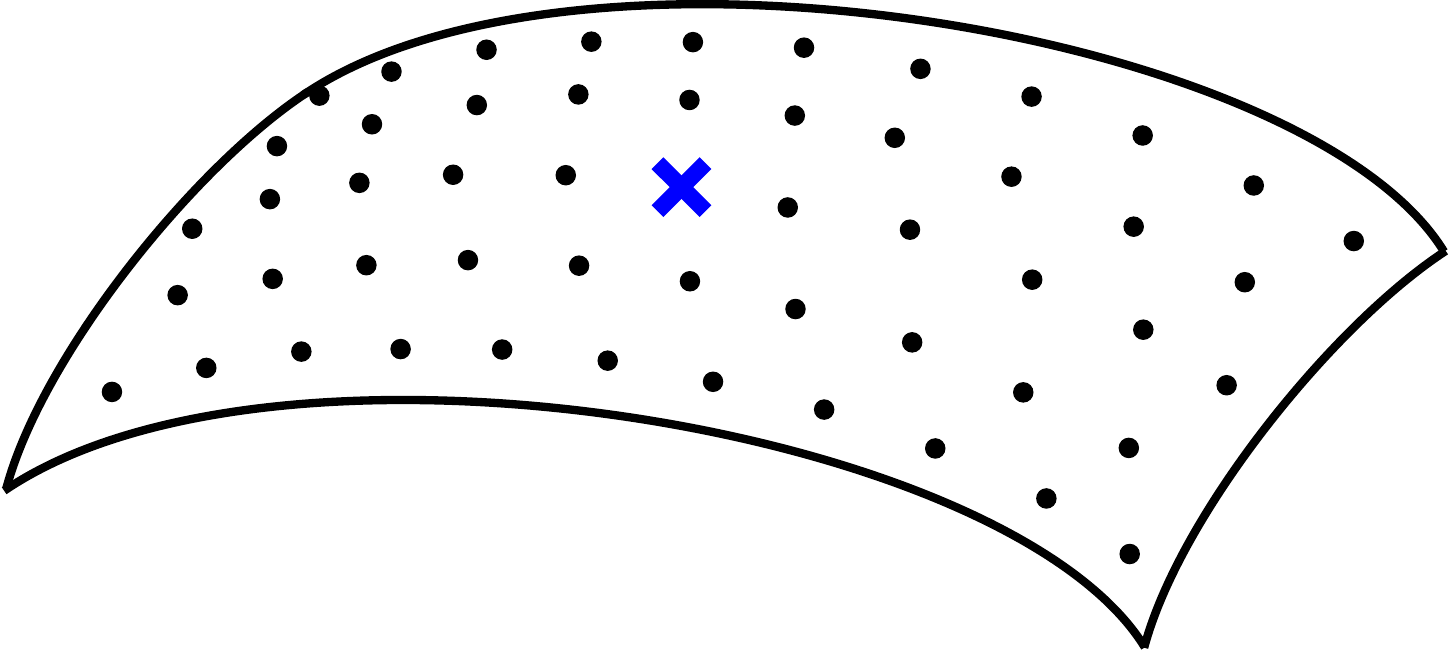}
\includegraphics{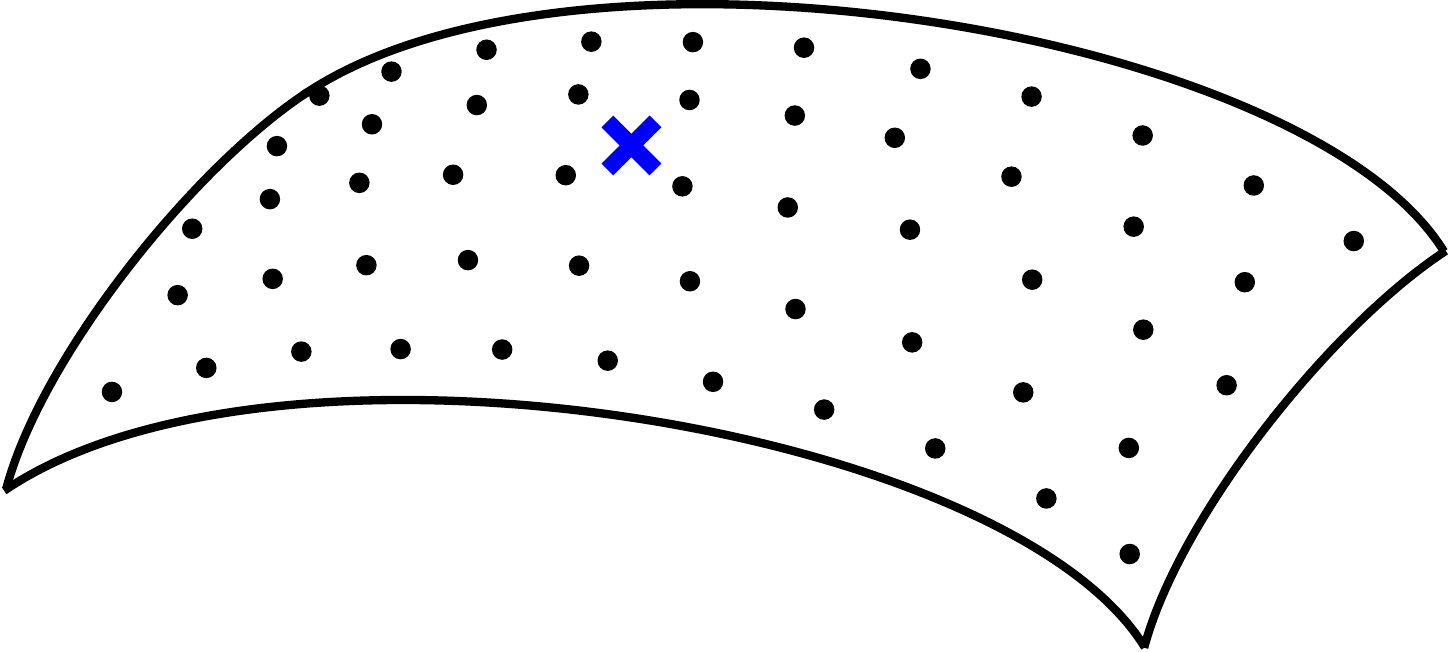}}}
\caption{A visual representation showing the two types of points of
$\bankpatch$ that we can choose to reconstruct. The \emph{left} cartoon shows
an example point that is in $\mat{H}$, and thus in $\bankpatch$. The
\emph{right} cartoon shows an example point that is not part of \mat{H} but is
in $\bankpatch$.}
\label{fig:cartoon}
\end{figure}

We test this embedding of the signal manifold to see how well various points in
the manifold can be reconstructed. The tests points we reconstruct are of two
types: 1) those from the original signal matrix $\mat{H}$, 2) those absent from
$\mat{H}$ but within $\bankpatch$. These two types of tests are illustrated in
Fig.~\ref{fig:cartoon}.

\begin{figure}
\center{\resizebox{\linewidth}{!}{\includegraphics{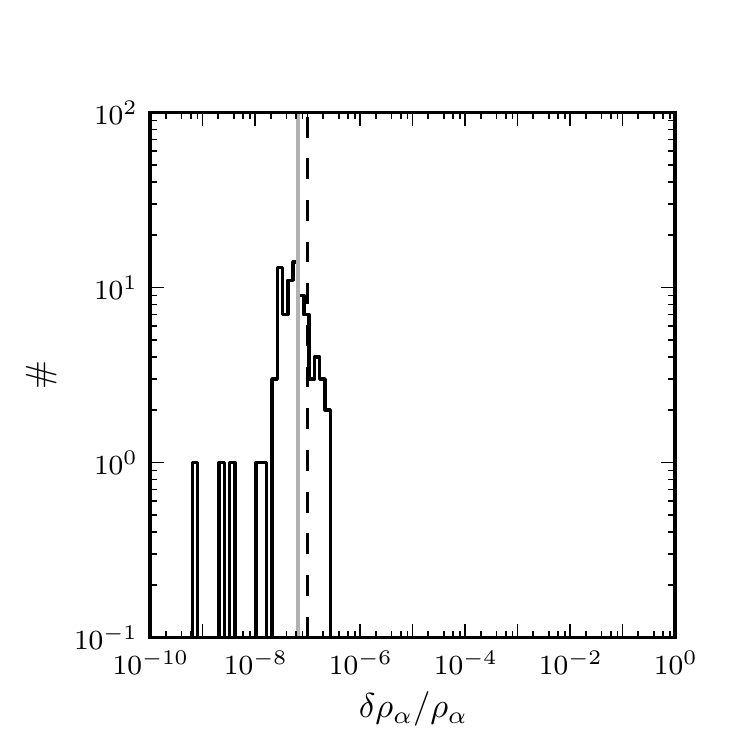}}}
\caption{A histogram of the reconstruction accuracy of the waveforms that went
into the construction of $\mat{H}$. As expected, the average reconstruction
(solid line) accuracy matches the expected fractional \ac{SNR} loss (dashed
line).}
\label{fig:reconH}
\end{figure}

A test of the first type is shown in Fig.~\ref{fig:reconH}. This shows that the
average reconstruction accuracy for points from $\mat{H}$ agrees with our
chosen value of $10^{-7}$. This result is expected as it is an extension of the
investigation from Fig.~4 of \cite{Cannon2010} applied to a more stringent
reconstruction accuracy.

\begin{figure*}
\center{\resizebox{\linewidth}{!}{\includegraphics{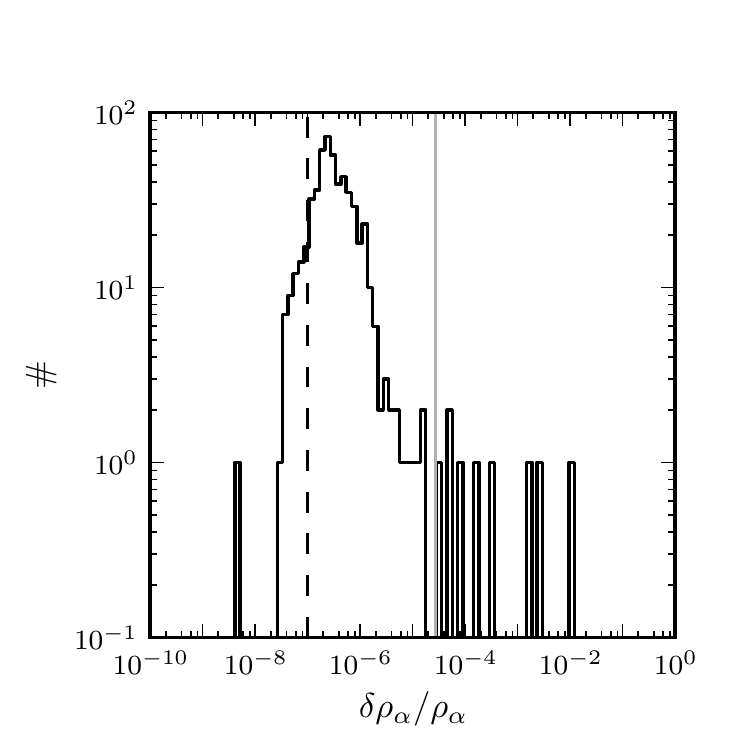}
\includegraphics{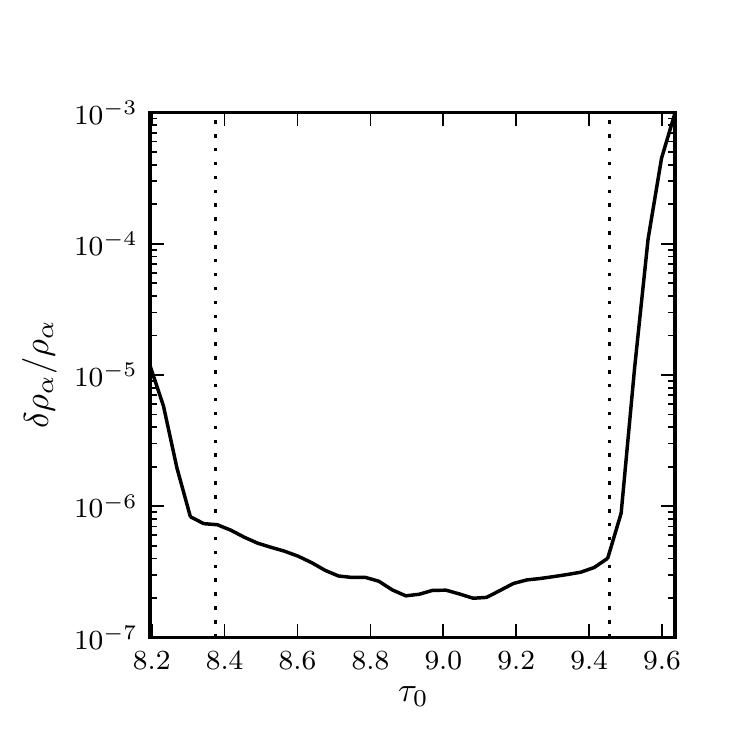}}}
\caption{(\emph{left}) A histogram of the mismatch between waveforms from
$\bankpatch$ that were not in $\mat{H}$ and their reconstructed versions.  The
peak of the mismatch is slightly above the expected fractional \ac{SNR} loss
for waveforms from $\mat{H}$ (dashed line). (\emph{right}) How the mismatch
varies across $\bankpatch$, averaged across the $\tau_3$ direction.  The
largest mismatches come from near the borders of the template bank in the
$\tau_0$ direction. Fig.~\ref{fig:reconP'} restricts our attention to the
central $75\%$ of the domain of $\bankpatch$, whose boundaries are shown as
vertical dotted lines.}
\label{fig:reconP}
\end{figure*}

A test of the second type is shown in Fig.~\ref{fig:reconP}. To choose points
uniformly from $\bankpatch$ but absent from $\mat{H}$, we generate a denser
template bank within the same region of parameter space described in the
Sec.~\ref{sec:embed}.  Specifically, we generate this template bank with a
minimal match of $99\%$. In order to test the reconstruction accuracy of these
waveforms, we project the real and imaginary parts of the waveforms onto the
basis vectors from the \ac{SVD} of $\mat{H}$
\begin{equation} v'_{\alpha \mu} = \frac{1}{\sigma_\mu} \sum_j h_{\alpha j}
u_{\mu j}, \end{equation}
where $v'_{\alpha \mu}$ represents a reconstruction coefficient associated with
the $\mu$th basis vector for the real or imaginary part of the $\alpha$th
waveform from the denser template bank, $\sigma_\mu$ is the $\mu$th element of
$\mat{\Sigma}$, $h_{\alpha j}$ is the $j$th time sample of the real or
imaginary part of the $\alpha$th waveform from the denser template bank, and
$u_{\mu j}$ is the $j$th time sample from the $\mu$th basis vector in
$\mat{U}$. The real and imaginary parts of the waveforms from the denser
template bank are then reconstructed using
\begin{equation} h'_{\alpha j} = \sum_\mu v'_{\alpha\mu} \sigma_{\mu} u_{\mu j}
\,.  \end{equation}

\begin{figure}
\center{\resizebox{\linewidth}{!}{\includegraphics{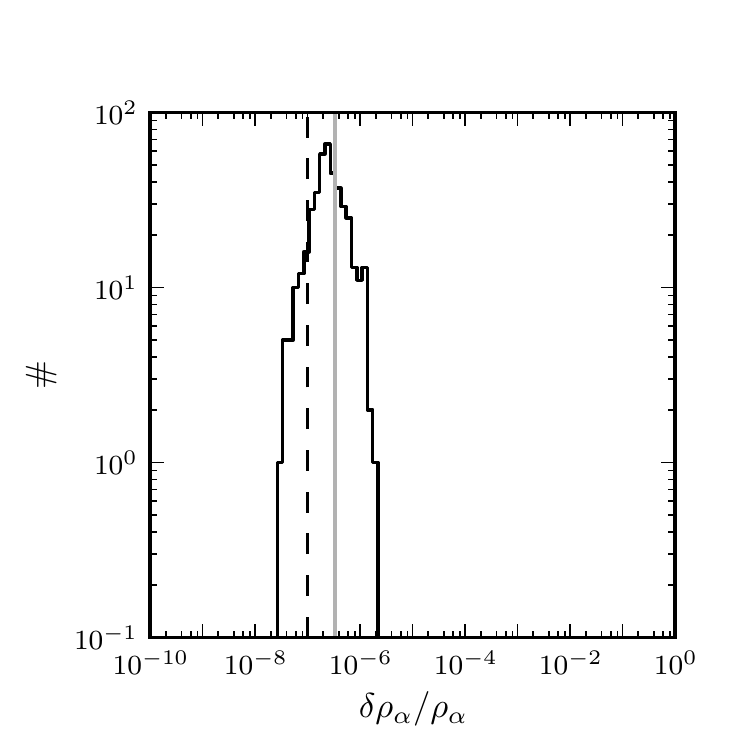}}}
\caption{Similar plots as in Fig.~\ref{fig:reconP}, eliminating test points
near the boundaries in the $\tau_0$ direction. (\emph{left}) A histogram of the
mismatch between waveforms from $\bankpatch$ that were not in $\mat{H}$ and
their reconstructed versions. The average reconstruction (solid line) accuracy
is slightly worse than the expected fractional \ac{SNR} loss (dashed line).}
\label{fig:reconP'}
\end{figure}

The distribution of $\delta\rho_\alpha / \rho_\alpha$ for these waveforms, left
panel of Fig.~\ref{fig:reconP}, shows a tail extending to large mismatches.
Examining where these large mismatches are located in parameter space, we find
they originate from near the boundaries of $\bankpatch$.  Removing the test
points near the boundaries in the $\tau_0$ direction, shown in
Fig.~\ref{fig:reconP'}, we find the tail of large mismatches disappears.

\begin{figure*}
\center{\resizebox{\linewidth}{!}{\includegraphics{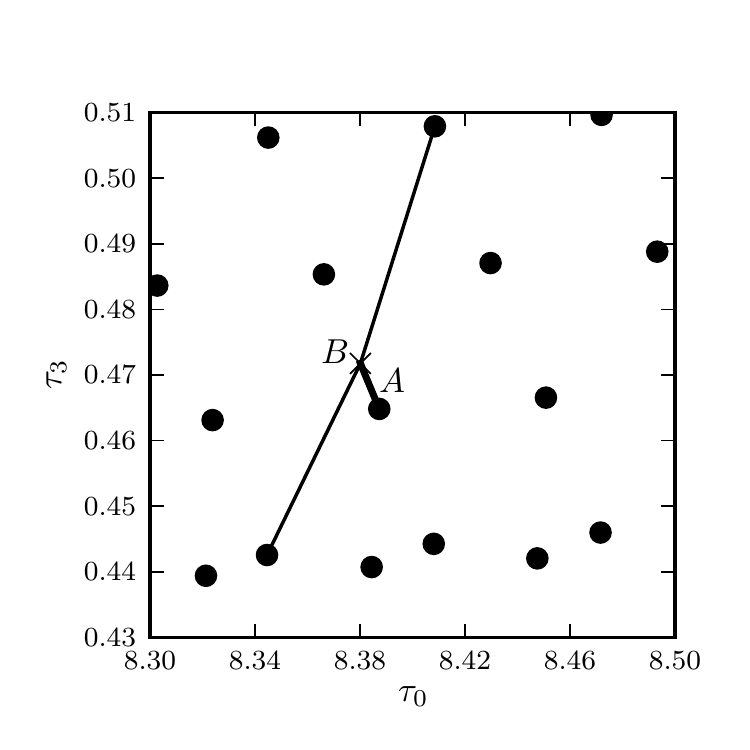}
\includegraphics{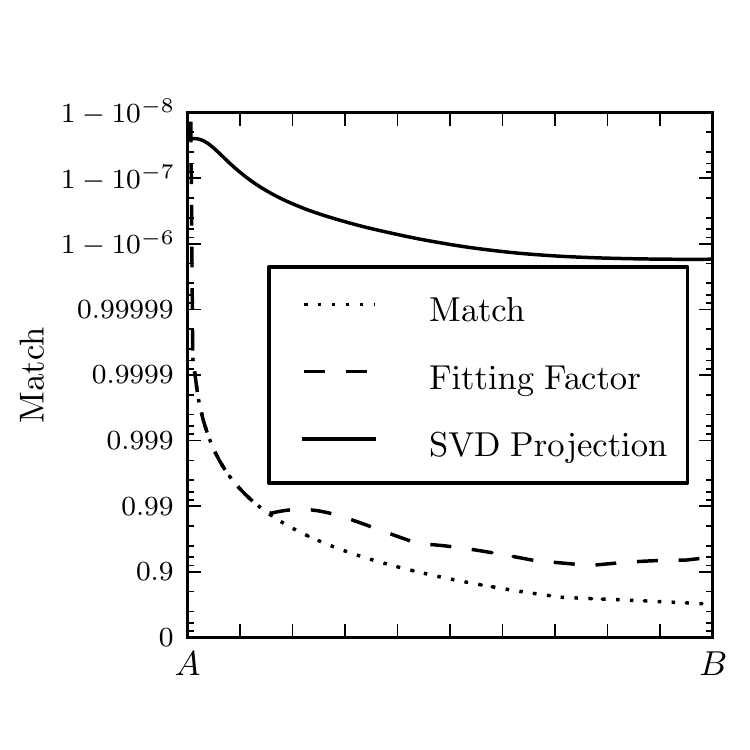}}}
\caption{(\emph{left}) A plot showing a smaller region of $\mathcal{P}$. The
circles are points whose waveforms go into $\mat{H}$. The line segment
$\overline{AB}$ connects one of those points, $A$, with the point $B$, the
central point of $A$ and two of its nearest neighbors. Point $B$ is situated
such that it should have a $89.9\%$ fitting factor with each of the surrounding
points.  (\emph{right}) How well a waveform from a given point of
$\overline{AB}$, $\vec{h}_p$, can be ``matched".  The Match shows the
normalized inner-product between $\vec{h}_A$ and the waveform from the
corresponding point along $\overline{AB}$.  The Fitting Factor shows the same
match maximized over phase and time. Since we are using a $89.9\%$ minimal match
bank, it is expected that the fitting factor falls to around that value. The
\ac{SVD} Projection is given by $1 - \delta\rho_\alpha / \rho_\alpha$. The
\ac{SVD} basis vectors are able to reconstruct to high accuracy all points
along the line.}
\label{fig:reconline}
\end{figure*}

An additional test of the second type, which systematically explores the
reconstruction accuracy near a point whose waveform went into $\mat{H}$, is
shown in Fig.~\ref{fig:reconline}. The left panel shows a set of three
nearest-neighbor templates. We investigate how the reconstruction accuracy
varies as one moves from point $A$ to the central point, point $B$. Point $B$
is assumed to have the largest mismatch between its waveform and the waveforms
from any of the three surrounding points. We also compute the mismatch between
the waveforms along $\overline{AB}$ and the waveform of point $A$ with and
without maximizing over phase and time, called the \emph{fitting factor} and
\emph{match} respectively. The fitting factor falls to the minimal match of the
template bank when comparing the waveforms from $A$ and $B$, which is expected
as the minimal match involves maximizing over phase and time. The
reconstruction accuracy associated with \ac{SVD} projection is consistently
high and close to the chosen reconstruction accuracy of 1 part in $10^7$.

\section{Discussion}
\label{sec:disc}

These investigations show that the \ac{SVD} can be used to find a set of basis
vectors that not only span the signal matrix $\mat{H}$, but also enclose the
signal manifold $\bankpatch$ sampled by $\mat{H}$.

\ac{GW} pipelines that search for known waveforms, such as \ac{GW}s from
\ac{CBC}s, commonly compute waveform consistency statistics that compare the
observed response of a template waveform filter to the data with what one would
expect given the presence of that signal. These consistency statistics are
found to perform better when the mismatch between the template waveform and the
signal waveform is small~\cite{Allen:2004}. Filtering with a fixed density
template bank can introduce mismatch between the nearest template and the
signal. This mismatch can be greatly reduced if one is able to find the exact
point in parameter space where the signal is located and filter the data using
that point. Using the \ac{SVD} basis vectors, one could reconstruct a point
closer to the point of the signal and improve the waveform consistency
statistics.

Parameter estimation techniques for \ac{GW}s from \ac{CBC}s often use Monte
Carlo Markov Chain algorithms to search the parameter space. This involves
producing waveforms and filtering the data against many points of the parameter
space. The \ac{SVD} could also be used to interpolate waveforms that are
expensive to compute, as in the case of waveforms produced by solving
differential equations. Also, if one filtered the data using the basis vectors
from the \ac{SVD}, it would be very easy to reconstruct to high accuracy the
output one would see if one had filtered the data using a waveform from
anywhere within the parameter space.

In order to gain benefit from these applications, it would be necessary to
determine the reconstruction coefficients in a computationally efficient
manner. This work has not tried to address this problem as: 1) it has assumed
the target waveforms are known, and 2) it computes the reconstruction
coefficients using computationally expensive inner products. Generation of
these reconstruction coefficients warrants future investigation as the benefits
derived from this technique would be substantial.

\acknowledgments

The authors would like to acknowledge the support of the LIGO Lab, NSF grants
PHY-0653653 and PHY-0601459, and the David and Barbara Groce Fund at Caltech.
LIGO was constructed by the California Institute of Technology and
Massachusetts Institute of Technology with funding from the National Science
Foundation and operates under cooperative agreement PHY-0757058.  Research at
Perimeter Institute is supported through Industry Canada and by the Province of
Ontario through the Ministry of Research \& Innovation. KC was supported by the
National Science and Engineering Research Council, Canada.  DK was supported in
part from the Max Planck Gesellschaft.  This work has LIGO document number
{LIGO-P1000039-v2}.

\bibliography{references}
\end{document}